\documentstyle[aps,prd]{revtex}

\begin{document}
\draft
\title{Disclination in Lorentz Space-Time}
\author{Sheng Li\thanks{%
E-mail: lisheng@itp.ac.cn}}
\address{Department of Applied Physics, Shanghai Jiaotong University, Shanghai
200030, P. R. China\\ Institute of Theoretical of Physics,
Academia Sinica, PO Box 2735, Beijing 100080, P. R. China\\
CCAST World Laboratory, Academia Sinica, Beijing 100080, P.
R. of China} \maketitle

\begin{abstract}
The disclination in Lorentz space-time is studied in detail by means of
topological properties of $\phi $-mapping. It is found the space-time
disclination can be described in term of a Dirac spinor. The size of the
disclination, which is proved to be the difference of two sets of $su(2)$%
-like monopoles expressed by two mixed spinors, is quantized topologically
in terms of topological invariants$-$winding number. The projection of
space-time disclination density along an antisymmetric tensor field is
characterized by Brouwer degree and Hopf index.
\end{abstract}

\pacs{PACS Numbers: 04.20.GZ, 02.40.-k, 11.15.-q}


\section{Introduction}

The topological defects certainly offer interesting arena
for the study of astrophysics and cosmology. These are
objects that may have formed during phase transitions in
the cooling down of the Early-Universe and may have played
a key role in the formation of the large scale structure of
the universe mainly through their gravitational
interaction\cite{Vilenkin,Brand}.

As a kind of topological defect, the disclination is caused by inserting a
solid angle into the flat space-time, which is described by the Riemannian
curvature $R_{\mu \nu \sigma }^\lambda $ or $SO(4)$ $(SO(3,1)$ in Lorentz
space-time$)$ gauge field tensor 2-form $F^{ab}$
\begin{equation}
\theta ^{ab}=\frac 12R_{\mu \nu \sigma }^\lambda e_\lambda ^ae^{\sigma
b}dx^\mu \wedge dx^\nu =-F^{ab}  \label{defination-disclination}
\end{equation}
in which $\theta ^{ab}$ is the disclination density. The size of the
disclination can be represented by the means of the surface integral of the
projection of the disclination density along an antisymmetric tensor field $%
\phi ^{ab}$
\begin{equation}
\Omega =\oint_\Sigma \theta ^{ab}\phi ^{ab}=-\oint_\Sigma F^{ab}\phi ^{ab}
\label{4w}
\end{equation}
where $\Sigma \,$ is a closed surface including the disclinations. The new
quantity $\Omega $ defined by (\ref{4w}) is dimensionless. In Riemann-Cartan
geometry, this effect is showed by the integral of the affine curvature
along a closed surface. Duan, Duan and Zhang\cite{Duan-Duan} had discussed
the disclinations in deformable material media by applying the gauge field
theory and decomposition theory of gauge potential. In their works, the
projection of disclination density along the gauge parallel vector was found
corresponding to a set of isolated disclinations in the three dimensional
sense and being topologically quantized.

In previous publication\cite{Duan-Li}, we studied the topological structure
of disclination in the 4-dimensional space-time of Euclidean signature. With
suitable coordinates, the projection of disclination density along a gauge
parallel tensor was found to be null everywhere except some singular points
where it has three dimensional $\delta $-function singularities. In fact,
just like the magnetic monopole theory, the topology of disclination
projection is described by the $U(1)$ projection of the gauge group. For $%
so(4)\cong su(2)\otimes su(2)$, the task is turned to find the topology of
the subgroup $U(1)\otimes U(1).$

However, in physics we are more interested in the space-time of Lorentzian
signature. There is no way in extending the methods to the case of
indefinite metric, because the Lorentz group $SO(3,1)$ is not, like $SO(4)$,
decomposable into a product of $SU(2)$ groups. In the present work, we find
a proper way to obtain the $U(1)$ projection of the Lorentz group by making
using of spinors. $\delta $-function singularities of the projection of the
disclination density are found again. The size of the space-time
disclination (\ref{4w}) is found to be a difference of two sets of solid
angles rather than a sum. The topological structure of the disclination is
discussed in detail.

\section{The gauge theory of Lorentz group and disclination projection}

In this paper, we use $a=1,2,3,4$\ to denote the Lorentz script with the
fourth component be pure imaginary and the signature of the metric of
Lorentz group is $(++++)$.

Let the four-dimensional Dirac matrix $\gamma _a$ are the basises of
Clifford algebra which satisfies
\begin{equation}
\gamma _a\gamma _b+\gamma _b\gamma _a=2\delta _{ab}.
\end{equation}
The antisymmetric tensor field $\phi ^{ab}$ on ${\bf M}$ can be expressed in
the following matrix form
\[
\phi =\frac 12\phi ^{ab}I_{ab}.
\]
in which $I_{ab}$ are the generators of the group $SO(3,1)$%
\begin{equation}
I_{ab}=\frac 14[\gamma _a,\gamma _b].
\end{equation}
The covariant derivative 1-form of $\phi $ is given by
\[
D\phi =d\phi -[\omega ,\phi ],
\]
where $\omega $ is the spin connection (gauge potential) which is also Lie
algebra-valued
\begin{equation}
\omega =\frac 12\omega ^{ab}I_{ab}.
\end{equation}
The curvature 2-form (gauge field) is
\begin{equation}
F=d\omega -\omega \wedge \omega .
\end{equation}

Take the chiral representation
\begin{equation}
\gamma _i=\left(
\begin{array}{cc}
0 & i\sigma _i \\
-i\sigma _i & 0
\end{array}
\right) \quad \quad \gamma _4=\left(
\begin{array}{cc}
0 & 1 \\
1 & 0
\end{array}
\right) .
\end{equation}
The generators $I_{ab}$ are
\begin{equation}
I_{ij}=\frac i2\varepsilon _{ijk}\left(
\begin{array}{cc}
\sigma _k & 0 \\
0 & \sigma _k
\end{array}
\right) \quad \quad I_{i4}=\frac i2\left(
\begin{array}{cc}
\sigma _k & 0 \\
0 & -\sigma _k
\end{array}
\right) ,
\end{equation}
in which $i,j=1,2,3.$ In the chiral representation the gauge field is
presented as
\begin{eqnarray}
F &=&\frac 12F^{ab}I_{ab}  \nonumber \\
&=&\frac i2\left(
\begin{array}{cc}
\frac 12F^{ij}\varepsilon _{ijk}\sigma _k+F^{i4}\sigma _i &  \\
& \frac 12F^{ij}\varepsilon _{ijk}\sigma _k-2F^{i4}\sigma _i
\end{array}
\right)  \nonumber \\
&=&\left(
\begin{array}{cc}
F_R &  \\
& F_L
\end{array}
\right) ,
\end{eqnarray}
where $F_R$ and $F_L$ are defined as
\begin{equation}
F_R=\frac i2(\frac 12\varepsilon ^{ijk}F^{jk}+F^{i4})\sigma _i\quad \quad
F_L=\frac i2(\frac 12\varepsilon ^{ijk}F^{jk}-F^{i4})\sigma _i.
\end{equation}
Also we can define $\omega _R$ and $\omega _L$ as
\begin{equation}
\omega _R=\frac i2(\frac 12\varepsilon ^{ijk}\omega ^{jk}+\omega
^{i4})\sigma _i\quad \quad \omega _L=\frac i2(\frac 12\varepsilon
^{ijk}\omega ^{jk}-\omega ^{i4})\sigma _i.
\end{equation}
It can be proved that
\begin{equation}
F_R=d\omega _R-\omega _R\wedge \omega _R\quad \quad F_L=d\omega _L-\omega
_L\wedge \omega _L.
\end{equation}
There exist the relations
\begin{equation}
F_R^{\dagger }=-F_L\quad \quad \omega _R^{\dagger }=-\omega _L.
\end{equation}

A Dirac spinor $\psi $ can be written as
\begin{equation}
\psi =\left(
\begin{array}{c}
\chi _R \\
\chi _L
\end{array}
\right) ,
\end{equation}
where $\chi _R$ and $\chi _L$ are the right-handed Weyl spinor and
left-handed Weyl spinor respectively. The Dirac spinor $\psi $ transforms
under $S\in SO(3,1)$ as
\begin{equation}
\psi \rightarrow S\psi .
\end{equation}
The covariant derivative of $\psi $ is
\begin{equation}
D\psi =d\psi -\omega \psi .
\end{equation}
For the Dirac conjugate spinor
\begin{equation}
\bar{\psi}=\psi ^{\dagger }\gamma _4=(\chi _L^{\dagger },\chi _R^{\dagger }),
\end{equation}
the covariant derivative is
\begin{equation}
D\bar{\psi}=d\bar{\psi}+\bar{\psi}\omega .
\end{equation}
Then the covariant derivatives of $\chi _{R,L}$ and $\chi _{R,L}^{\dagger }$
are
\begin{eqnarray}
D\chi _R &=&d\chi _R-\omega _R\chi _R\quad \quad D\chi _L=d\chi _L-\omega
_L\chi _L \\
D\chi _R^{\dagger } &=&d\chi _R^{\dagger }+\chi _R^{\dagger }\omega _R\quad
\quad D\chi _L^{\dagger }=d\chi _L^{\dagger }+\chi _L^{\dagger }\omega _L.
\end{eqnarray}

Define an antisymmetric tensor $\phi ^{ab}$ as
\begin{equation}
\phi ^{ab}=-\frac i2\bar{\psi}I_{ab}\psi
\end{equation}
It is easy to prove $\phi ^{ab}$ is a $SO(3,1)$ covariant tensor with $\phi
^{ij}$ $(i,j=1,2,3)$ are real and $\phi ^{i4}$ are pure imaginary. Then the
projection of curvature 2-form $F^{ab}$ on $\phi ^{ab}$ is
\begin{equation}
F^{ab}\phi ^{ab}=-i\bar{\psi}F\psi
\end{equation}
Therefore, we get the disclination projection density in Lorentz Space-time
as
\begin{equation}
F_p=-F^{ab}\phi ^{ab}=i\bar{\psi}F\psi  \label{discl}
\end{equation}
which is a Lorentz invariant.

Noticing that
\begin{equation}
F\psi =-D^2\psi ,
\end{equation}
we get
\begin{equation}
-i\bar{\psi}F\psi =iD(\bar{\psi}D\psi )-iD\bar{\psi}\wedge D\psi .  \label{1}
\end{equation}
For $\bar{\psi}D\psi $ is a Lorentz scalar, we have
\begin{eqnarray}
iD(\bar{\psi}D\psi ) &=&id(\bar{\psi}D\psi )  \nonumber \\
&=&id\bar{\psi}\wedge d\psi -id(\bar{\psi}\omega \psi ).  \label{2}
\end{eqnarray}
Denote the projection of the connection 1-form $\omega ^{ab}$ on $\phi ^{ab}$
as
\begin{equation}
A=\omega ^{ab}\phi ^{ab}=-i\bar{\psi}\omega \psi .  \label{3}
\end{equation}
Substituting equations (\ref{1}), (\ref{2}) and (\ref{3}) into the
disclination projection density (\ref{discl}), we get
\begin{eqnarray}
F_p &=&-id\bar{\psi}\wedge d\psi -dA+iD\bar{\psi}\wedge D\psi  \nonumber \\
&=&-id\chi _L^{\dagger }\wedge d\chi _R-id\chi _R^{\dagger }\wedge d\chi
_L-dA+iD\bar{\psi}\wedge D\psi .
\end{eqnarray}
Define two mixed spinors as
\begin{eqnarray}
\xi _{+} &=&\frac 12(\chi _R+\chi _L) \\
\xi _{-} &=&\frac 12(\chi _R-\chi _L).
\end{eqnarray}
We get the disclination projection density in terms of the mixed spinors
\begin{equation}
F_p=-2id\xi _{+}^{\dagger }\wedge d\xi _{+}+2id\xi _{-}^{\dagger }\wedge
d\xi _{-}+dA+iD\bar{\psi}\wedge D\psi .  \label{projectionfinal}
\end{equation}
Choose proper $\psi $ such that
\[
\psi ^{\dagger }\psi =4\quad and\quad \bar{\psi}\psi =0,
\]
which makes $\xi _{+}$ and $\xi _{-}$ satisfy
\begin{equation}
\xi _{+}^{\dagger }\xi _{+}=\xi _{-}^{\dagger }\xi _{-}=1.
\end{equation}
For the nontrivial topology of space-time, there must exist some points
where the mixed spinors $\xi _{\pm }$ are singular. Therefore $F_p$ is not a
total derivative globally. For the mixed spinors do not transform
covariantly under the Lorentz transformation, the first term and the second
term of the disclination projection density (\ref{projectionfinal}) do not
remain invariant under the Lorentz transformation respectively. However, the
total disclinations projection density $F_d$ do hold invariant.

Define two vectors $n_{\pm }^i$ as
\begin{equation}
n_{\pm }^i=\xi _{\pm }^{\dagger }\sigma ^i\xi _{\pm },  \label{unit}
\end{equation}
which make $n_{\pm }^i$ be unit vectors
\begin{equation}
n_{\pm }^in_{\pm }^i=1.
\end{equation}
Substitute (\ref{unit}) into the first and the second terms of the right
hand side of equation (\ref{projectionfinal}), it can be proved
\begin{equation}
-2id\xi _{\pm }^{\dagger }\wedge d\xi _{\pm }=\frac 12\varepsilon
^{ijk}n_{\pm }^idn_{\pm }^j\wedge dn_{\pm }^k.  \label{monopole}
\end{equation}
The definition (\ref{unit}) actually gives a Hopf map
\begin{equation}
S^3\rightarrow S^2
\end{equation}
and equation (\ref{monopole}) gives two $su(2)$-like monopoles\cite
{Ryder,minami}. If $\psi $ is taken as a gauge parallel spinor
\begin{equation}
D\psi =0,
\end{equation}
the disclination projection density becomes
\begin{equation}
F_p=\frac 12\varepsilon ^{ijk}n_{+}^idn_{+}^j\wedge dn_{+}^k-\frac 12%
\varepsilon ^{ijk}n_{-}^idn_{-}^j\wedge dn_{-}^k+dA.
\end{equation}
Then the size of the space-time disclination is
\begin{equation}
\Omega =\oint_\Sigma \frac 12\varepsilon ^{ijk}n_{+}^idn_{+}^j\wedge
dn_{+}^k-\frac 12\varepsilon ^{ijk}n_{-}^idn_{-}^j\wedge dn_{-}^k+dA.
\label{size-1}
\end{equation}
For the surface $\Sigma $ is closed, the third term of the right hand side
of (\ref{size-1}) contribute nothing to $\Omega $, i.e.
\begin{equation}
\Omega =\oint_\Sigma \frac 12\varepsilon ^{ijk}n_{+}^idn_{+}^j\wedge
dn_{+}^k-\frac 12\varepsilon ^{ijk}n_{-}^idn_{-}^j\wedge dn_{-}^k.
\end{equation}
Using Stokes formula, $\Omega $ can be expressed as
\begin{equation}
\Omega =\int_V\frac 12\varepsilon ^{ijk}dn_{+}^i\wedge dn_{+}^j\wedge
dn_{+}^k-\frac 12\varepsilon ^{ijk}dn_{-}^i\wedge dn_{-}^j\wedge dn_{-}^k,
\end{equation}
in which $\partial V=\Sigma $. Let us choose coordinates $y=\left(
u^1,u^2,u^3,v\right) $ on ${\bf M}$ such that $u=\left( u^1,u^2,u^3\right) $
are intrinsic coordinate on $V$. For the coordinate component $v$ does not
belong to $V$, $\Omega $ becomes
\begin{eqnarray*}
\Omega &=&\int_V(\frac 12\varepsilon ^{ijk}\partial _\alpha n_{+}^i\partial
_\beta n_{+}^j\partial _\gamma n_{+}^k-\frac 12\varepsilon ^{ijk}\partial
_\alpha n_{-}^i\partial _\beta n_{-}^j\partial _\gamma n_{-}^k)du^\alpha
\wedge du^\beta \wedge du^\gamma \\
&=&\int_V(\frac 12\varepsilon ^{ijk}\varepsilon ^{\alpha \beta \gamma
}\partial _\alpha n_{+}^i\partial _\beta n_{+}^j\partial _\gamma n_{+}^k-%
\frac 12\varepsilon ^{ijk}\varepsilon ^{\alpha \beta \gamma }\partial
_\alpha n_{-}^i\partial _\beta n_{-}^j\partial _\gamma n_{-}^k)d^3u,
\end{eqnarray*}
where $\alpha ,\beta ,\gamma =1,2,3$ and $\partial _\alpha =\partial
/\partial u^\alpha $. Define solid angle densities as
\begin{equation}
\rho =\frac 12\varepsilon ^{ijk}\varepsilon ^{\alpha \beta \gamma }\partial
_\alpha n_{+}^i\partial _\beta n_{+}^j\partial _\gamma n_{+}^k-\frac 12%
\varepsilon ^{ijk}\varepsilon ^{\alpha \beta \gamma }\partial _\alpha
n_{-}^i\partial _\beta n_{-}^j\partial _\gamma n_{-}^k.
\label{solid-density}
\end{equation}
Then we get
\begin{equation}
\Omega =\int_V\rho d^3u.
\end{equation}

\section{Topological structure of the space-time disclination}

Now one see the final equation (\ref{solid-density}) is similar to that of
space-time of Euclidean signature in the previous paper\cite{Duan-Li} except
the disclination size is a difference rather than a sum.

The unit vectors $n_{\pm }^i$, can be express as
\begin{equation}
n_{\pm }^i=\frac{\phi _{\pm }^i}{||\phi _{\pm }||}
\end{equation}
where $\phi _{\pm }^i$, are smooth vectors along the direction of $n_{\pm
}^i $ respectively. Substituting it into (\ref{solid-density}), and using
the Laplacian relation we can obtain the topological structure of the
disclination in Lorentz space-time
\begin{equation}
\rho =4\pi \delta ^3(\phi _{+})J(\frac{\phi _{+}}u)-4\pi \delta ^3(\phi
_{-})J(\frac{\phi _{-}}u)  \label{density-1}
\end{equation}
and
\begin{equation}
\Omega =4\pi \int_V(\delta ^3(\phi _{+})J(\frac{\phi _{+}}u)-\delta ^3(\phi
_{-})J(\frac{\phi _{-}}u))d^3u.
\end{equation}
in which the Jacobian $J(\frac{\phi _{\pm }}u)$ are
\begin{equation}
\varepsilon ^{ijk}J(\frac{\phi _{\pm }}u)=\varepsilon ^{\alpha \beta \gamma
}\partial _\alpha \phi _{\pm }^i\partial _\beta \phi _{\pm }^j\partial
_\gamma \phi _{\pm }^k.
\end{equation}

Suppose that $\phi _{\pm }^i(x)$, possess $K_{\pm }$ isolated zeros
respectively, the solutions of $\phi _{\pm }(u^1,u^2,u^3,v)=0$ can be
expressed in terms of $u=(u^1,u^2,u^3)$ as
\begin{equation}
u^i=z^i(v),\quad \quad \quad \quad \quad i=1,2,3
\end{equation}
and
\begin{equation}
\phi ^i(z_l^1(v),z_l^2(v),z_l^3(v),v)\equiv 0,
\end{equation}
where the subscript $l=1,2,\cdots ,K$ represents the $l$th zero of $\phi ^i$%
, i.e.
\begin{equation}
\phi _{\pm }^i(z_l^i)=0,\quad \quad \quad l=1,2,\cdots ,K_{\pm };\quad
A=1,2,3.
\end{equation}
It is easy to get the following formula from the ordinary theory of $\delta $%
-function that
\begin{equation}
\delta ^3(\phi _{\pm })J(\frac{\phi _{\pm }}u)=\sum_{i=1}^{K_{\pm }}\beta
_{\pm l}\eta _{\pm l}\delta ^3(u-z_{\pm l}),
\end{equation}
in which
\begin{equation}
\eta _{\pm l}=signJ(\frac{\phi _{\pm }}u)|_{x=z_{\pm l}}=\pm 1
\end{equation}
is the Brouwer degree of $\phi $-mapping and $\beta _{\pm l}$ are positive
integers called the Hopf index of map $\phi _{\pm }$ which means while the
point $x$ covers the region neighboring the zero $x=z_{\pm l}$ once, $\phi
_{\pm }$ covers the corresponding region $\beta _{\pm l}$ times. Therefore
the slid angle density becomes
\begin{equation}
\rho =4\pi \sum_{l=1}^{K_{+}}\beta _{+l}\eta _{+l}\delta ^3(u-z_{+l})-4\pi
\sum_{l=1}^{K_{-}}\beta _{-l}\eta _{-l}\delta ^3(u-z_{-l})  \label{density-d}
\end{equation}
and
\begin{equation}
\Omega =4\pi \sum_{l=1}^{K_{+}}\beta _{+l}\eta _{+l}-4\pi
\sum_{l=1}^{K_{-}}\beta _{-l}\eta _{-l}.  \label{finalsize}
\end{equation}
It is obvious from (\ref{finalsize}) that the size of the space-time
disclination is quantized for topological reason.

In fact the winding numbers $W_{\pm l}$ of the map $\phi _{\pm }$ around the
zeroes $z_{\pm l}$ are
\[
W_{\pm l}=\beta _{\pm l}\eta _{\pm l}
\]
Hence, the space-time disclinations is quantized by the winding numbers as
\begin{equation}
\Omega =4\pi \sum_{l=1}^{K_{+}}W_{+l}-4\pi \sum_{l=1}^{K_{-}}W_{-l}.
\end{equation}
or by the winding numbers of the map $\phi _{\pm }$ around the surface $%
\Sigma $
\begin{equation}
\Omega =4\pi W_{+}-4\pi W_{-}.  \label{windingnumber}
\end{equation}
The winding number $W_{\pm }$ of the surface $\Sigma $ can be interpreted
or, indeed, defined as the degree of the mappings $\phi _{\pm }$ onto $%
\Sigma $. Then the space-time disclination is
\begin{equation}
\Omega =4\pi \deg \phi _{+}-4\pi \deg \phi _{-}.
\end{equation}
We find that (\ref{density-d}) is the exact density of a system of $K_{+}$
and $K_{-}$ classical point-like objects with quantized ``charge'' $\beta
_{+l}\eta _{+l}$ and $\beta _{-l}\eta _{-l}$ in space-time, i.e. the
topological structure of disclinations formally corresponds to a point-like
system. These point objects may be called disclination points as in nematic
crystals\cite{Klem}. In \cite{Genn}, it was shown that the existence of
disclination points is related to a kind of broken symmetries. The
dislocations and disclinations appear as singularities of distortions of an
order parameter\cite{Frie}. In our paper, the disclination points are
identified with the isolated zero points of vector field $\phi _{\pm }^i(x)$%
. From (\ref{density-1}) we know that these singularities are those of the
disclination density as well.

\section{Conclusion}

In this paper, we have studied the topological structure, global and local
properties of disclinations in the general $4$-dimensional Lorentz
space-time. By defining an antisymmetric tensor in terms of a Dirac spinor,
we get the disclination projection density. The projection of disclination
is proved to be the difference of two sets of isolated disclinations, each
of which corresponds to a $su(2)$-like monopole expressed by some mixed
spinor. We showed that the size of space-time disclination is quantized
topologically. The positions of the disclinations are determined by the
zeroes of two mixed spinors. And the Hopf index and Brouwer degree classify
the disclinations and characterize the local nature of the space-time
disclinations. For this quantization of the size of the space-time
disclinations is related to the space-time curvature directly, it has close
relationship with the quantization of space-time.

\end{document}